\documentclass[pre,aps,twocolumn,showpacs,superscriptaddress,floatfix]{revtex4-2}

\usepackage{mathtools}
\usepackage[usenames,dvipsnames]{xcolor}
\usepackage{amsmath}
\usepackage{amssymb,mathrsfs}
\usepackage{graphicx}
\usepackage[colorlinks=true]{hyperref}
\usepackage{soul}
\usepackage{xcolor}

\newcommand{\prt}{\partial}
\newcommand{\al}{\alpha}

\newcommand{\eps}{\varepsilon}

\newcommand{\om}{\omega}

\newcommand{\bu}{\mathbf{u}}
\newcommand{\br}{\mathbf{r}}
\newcommand{\bv}{\mathbf{v}}
\newcommand{\bt}{\mathbf{t}}
\newcommand{\bn}{\mathbf{n}}

\begin{document}

\title{Contour dynamics of two-dimensional dark solitons}

\author{A. M. Kamchatnov}
 \affiliation{Institute of Spectroscopy, Russian Academy of Sciences, Troitsk, Moscow, 108840, Russia}

\begin{abstract}
Equations for contour dynamics of dark solitons are obtained for the general form of the nonlinearity
function. Their self-similar solution which describes the nonlinear stage of the bending instability
of dark solitons is studied in detail.
\end{abstract}



\maketitle

\section{Introduction}

Dynamics of dark solitons plays an important role in nonlinear optics and physics of
Bose-Einstein condensates (BECs) (see, e.g., \cite{ka-03,ps-03} and references therein).
In particular, if the condensate is confined in a quasi-1D harmonic trap, such a soliton
oscillates with the frequency different from the trap frequency on the contrary to the
behavior of bright solitons \cite{ba-00,kp-04,bkp-06,ks-09}. Dynamics of dark solitons
becomes even more complicated in 2D case which is typical, for example, in physics of
polariton condensates formed in planar microresonators (see, e.g., \cite{dngl-13}).
As was shown in Refs.~\cite{kp-70,zakh-75,kt-88}, 2D dark solitons are unstable with
respect to the bending (`snake') instability. As a result,  a dark soliton breaks down
with formation of vortices, and this phenomenon was observed experimentally in
Refs.~\cite{tcldk-96,msz-96,msaz-96,ahrfccc-01}.

Theoretical description of transition from the exponential growth of the unstable
``snake'' modes at the linear stage of their evolution to the nonlinear stage leading
eventually to formation of vortices is a difficult task and several possible scenarios
were identified depending on the solion's amplitude \cite{psk-95} (see also review article
\cite{kp-2000} and references therein). An interesting approach to description of nonlinear
evolution of instability of deep enough dark solitons
was suggested in Ref.~\cite{ms-10,mss-11} on the basis of the {\it contour dynamics} \cite{bkkl-84}.
Mironov {\it et al} \cite{ms-10,mss-11} assumed that the local radius of curvature of a dark soliton
is much greater than its local width, so that the position of this soliton can be represented
with high accuracy by a curved line---the soliton's `contour'. Then the bending dynamics
of such a contour is determined by two variables---the local velocity of the soliton
and its local curvature. Mironov {\it et al} \cite{ms-10,mss-11} derived the equations governing this
dynamics in framework of the perturbation theory for the case of BEC dynamics obeying
the standard Gross-Pitaevskii equation and studied the nonlinear stage of development of
instability of dark solitons. Later this theory was generalized in Ref.~\cite{ssok-14}
to the instability dynamics of dark solitons in polariton condensate. To avoid any confusion,
we would like to stress that the contour dynamics of Mironov {\it et al} \cite{ms-10,mss-11}
differs from dynamics of contours around 2D vortex patches developed by N.~J.~Zabusky 
{\it et al} \cite{zhr-79} (see also review article \cite{pullin-92} and references therein).

In this paper, we derive the equations of contour dynamics of dark solitons for media
whose evolution obeys the generalized Gross-Pitaevskii equation (or generalized nonlinear
Schr\"{o}dinger (NLS) equation)
\begin{equation}\label{eq1}
  i\psi_t+\frac12(\psi_{xx}+\psi_{yy})-f(|\psi|^2)\psi=0
\end{equation}
with general form of the positive nonlinearity function $f>0$. Our derivation is based
on physical reasoning rather than on the formal application of the perturbation theory.
After that we study analytically in some detail the self-similar solutions of these
equations. These solutions describe the nonlinear stage of the bending instability
of dark solitons and considerably extend the results obtained in Ref.~\cite{ms-10,mss-11}.

\section{Dark soliton solution of the generalized NLS equation}

First, we shall present here the basic results of the dark soliton theory. For definiteness,
we shall interpret Eq.~(\ref{eq1}) as the Gross-Pitaevskii equation for dynamics of BEC,
so that $\rho=|\psi|^2$ has the meaning of the condensate's density and the gradient of
the phase $\bu=\nabla\phi$ has the meaning of the condensate's flow velocity. These
definitions imply the representation of the condensate wave function $\psi$ in the form
\begin{equation}\label{eq2}
  \psi=\sqrt{\rho(\br,t)}\exp[i\phi(\br,t)-i\mu t],
\end{equation}
where
\begin{equation}\label{eq3}
  \mu=f(\rho_0)
\end{equation}
is the chemical potential of the condensate with a uniform density $\rho_0$ far from the dark
soliton. Substitution of Eq.~(\ref{eq2}) into Eq.~(\ref{eq1}) and standard calculations yield
the equations of BEC dynamics in the hydrodynamic-like form
\begin{equation}\label{eq4}
  \begin{split}
  & \rho_t+\nabla(\rho\bu)=0,\\
  & \bu_t+(\bu\nabla)\bu+\nabla f(\rho)+\nabla\left(\frac{(\nabla\rho)^2}{8\rho^2}-
  \frac{\Delta\rho}{4\rho}\right)=0.
  \end{split}
\end{equation}
Linearization of these equations with respect to a uniform quiescent BEC with $\rho=\rho_0$,
$\bu=\bu_0=0$ gives the Bogoliubov dispersion relation
\begin{equation}\label{eq5}
  \om=k\sqrt{c_0^2+k^2/4}
\end{equation}
for linear waves $\propto\exp[i(\mathbf{k}\cdot\br-\om t)]$, where $c_0$ is the sound
velocity
\begin{equation}\label{eq6}
  c_0=\sqrt{\rho_0f'(\rho_0)}
\end{equation}
of waves in the long wavelength limit.

It is not difficult to find the soliton solution of Eqs.~(\ref{eq4}) for which the
variables $\rho$ and $\bu$ depend only on the distance $\xi=(\bv/v)\cdot(\br-\bv t)$ from
the straight line normal to vector $\bv$ of the soliton velocity ($v=|\bv|$).
Substitution of the {\it ansatz} $\rho=\rho(\xi)$, $\bu=(\bv/v)u(\xi)$ gives with account
of the boundary conditions $\rho\to\rho_0$, $u\to0$ as $|\xi|\to\infty$ the relationship
\begin{equation}\label{eq7}
  u(\xi)=v\left(1-\frac{\rho_0}{\rho}\right)
\end{equation}
and the equation for $\rho(\xi)$ (see Ref.~\cite{ks-09})
\begin{equation}\label{eq8}
  \rho_{\xi}^2=Q(\rho),
\end{equation}
where
\begin{equation}\label{eq9}
  Q(\rho)=8\rho\int_{\rho}^{\rho_0}[f(\rho_0)-f(\rho')]d\rho'-4v^2(\rho_0-\rho)^2.
\end{equation}
Integration of Eq.~(\ref{eq8}) gives at once
\begin{equation}\label{eq10}
  \xi=\int_{\rho_m}^{\rho}\frac{d\rho}{\sqrt{Q(\rho)}},
\end{equation}
where $\rho_m$ is the minimal density at the center $\xi=0$ of the soliton.
The function $Q(\rho)$ has a double zero at $\rho=\rho_0$, hence
$\left.dQ/d\rho\right|_{\rho=\rho_0}=0$, and this equation yields the relationship
between $\rho_m$ and the soliton velocity $v$,
\begin{equation}\label{eq11}
  v^2=\frac{Q_0(\rho_m)}{4(\rho_0-\rho_m)^2},
\end{equation}
where
\begin{equation}\label{eq12}
  Q_0(\rho)=8\rho\int_{\rho}^{\rho_0}[f(\rho_0)-f(\rho')]d\rho'.
\end{equation}
Inverse of the function $\xi=\xi(\rho)$ defined by Eq.~(\ref{eq10}) gives the profile
$\rho=\rho(\xi)$ of density of the condensate with the soliton propagating through it,
and substitution of this expression for $\rho(\xi)$ into Eq.~(\ref{eq7}) provides 
the profile of the corresponding flow velocity $u=u(\xi)$.

Soliton's energy per unit length can be calculated by the method of Ref.~\cite{bkp-06}
and it is given by the expression (see Ref.~\cite{ks-09})
\begin{equation}\label{eq13}
  \eps=\frac12\int_{\rho_m}^{\rho_0}\frac{Q_0(\rho)d\rho}{\rho\sqrt{Q(\rho)}}.
\end{equation}
Here $\rho_m$ is a function of $v$ according to Eq.~(\ref{eq11}) and the same is true for
functions $Q_0(\rho)$ and $Q(\rho)$, so we can consider the soliton's energy as a known
function of its velocity $v$:
\begin{equation}\label{eq14}
  \eps=\eps(v).
\end{equation}
For example, in case of standard `Kerr-like' nonlinearity $f(\rho)=\rho$ our formula
reduces to the well-known expression
\begin{equation}\label{eq15}
  \eps(v)=\frac43(\rho_0-v^2)^{3/2}.
\end{equation}
In all above formulas the background condensate density $\rho_0$ is a constant parameter.

Now we can proceed to derivation of equations of the contour dynamics.

\section{Equations of contour dynamics}

We assume that the instant position of the 2D dark soliton in the $(x,y)$-plane is given
in a parametric form
\begin{equation}\label{eq16}
  \br(s)=(x(s),y(s)),
\end{equation}
where $s$ is the length of soliton's arc starting from some `zero' point to the point (\ref{eq16}).
Following the rules of elementary differential geometry (see, e.g., \cite{pressley}),
we introduce the tangent vector $\bt(s)=d\br/ds$, $|\bt|=1$, and the unit normal vector
$\bn(s)$, $|\bn|=1$, which obey the Frenet-Serret equations
\begin{equation}\label{eq17}
  \frac{\prt\bt}{\prt s}=\kappa\bn,\qquad \frac{\prt\bn}{\prt s}=-\kappa\bt
\end{equation}
for plane curves, where $\kappa$ is the curvature of the curve at the point $\br$.
We use the partial derivatives here to indicate they are taken for
an instant position of the curve (\ref{eq16}) (soliton's contour). Now we take into
account that the soliton moves and deforms, so that its point with the coordinate $s$ at
the moment of time $t$ has velocity
\begin{equation}\label{eq18}
  \frac{\prt\br}{\prt t}=v\bn+w\bt,
\end{equation}
where the first term corresponds to the motion of the curve in the normal direction
and the second term corresponds to its stretching with change of the length $s$.
The condition $\br_{ts}=\br_{st}$ yields
\begin{equation}\label{eq19}
  w_s=\kappa v,\qquad \bt_t=(v_s+\kappa w)\bn,
\end{equation}
that is
\begin{equation}\label{eq20}
  w=\int_0^s\kappa vds'.
\end{equation}
If we differentiate the second equation (\ref{eq19}) with respect to $s$ and replace
$\bt_s$ and $w$ with the use of (\ref{eq17}) and (\ref{eq20}), then we get
$$
\kappa_t\bn+\kappa \bn_t=\left(v_{ss}+\int_0^s\kappa vds'\cdot\kappa_s+\kappa^2v\right)\bn
-\kappa(v_s+\kappa w)\bt,
$$
that is, with account of $(\bn\cdot\bt)_t=0$ and the second Eq.~(\ref{eq19}), we obtain
\begin{equation}\label{eq21}
  \kappa_t-\int_0^s\kappa vds'\cdot\kappa_s=v_{ss}+\kappa^2v.
\end{equation}
This is a kinematic equation of the contour dynamics which follows from purely
geometric consideration (see also discussions of the contour dynamics in
Refs.~\cite{bkkl-84,ms-10,mss-11}). The second term in its left-hand side has the meaning of
change of the curvature $\kappa$ due to transfer of soliton's points along the arc with
velocity $w$: $ds=-wdt$. In other words, the contour's motion leads to the reparametrization
of its points and the derivative in the left-hand side of Eq.~(\ref{eq21}) is
interpreted as a `substantial derivative':
\begin{equation}\label{eq22}
  \frac{d\kappa}{dt}=\left(\frac{\prt}{\prt t}+\frac{ds}{dt}\frac{\prt}{\prt s}\right)\kappa=
  \left(\frac{\prt}{\prt t}-\int_0^s\kappa vds'\cdot\frac{\prt}{\prt s}\right)\kappa.
\end{equation}

Now we turn to derivation of the dynamical equation for the soliton's contour motion.
The energy of a dark soliton decreases with increase of its velocity. Actually, this is
the reason for its bending instability \cite{os-76,kp-08}. If a straight soliton moving along
the $x$-axis with velocity $v_0$ undergoes a small bending disturbance $x'(y,t)$, then its
instant position is described by the function $x=v_0t+x'(y,t)$ and its local velocity
\begin{equation}\label{eq23}
  v=v_0+x'_t(y,t)
\end{equation}
becomes a function of the coordinate $y$ along the soliton. For small $x'$ the velocity
$v$ can still be considered as the velocity component $v_n$ normal to the soliton's
contour at the point $y$: $v_n\approx v$. Then the energy per unit length is equal to
$\eps(v)$ and it is different from $\eps(v_0)$ due to the local disturbance, that is
due to the growth of the length $l$ with the rate $dl/dt=\kappa v_n\approx\kappa v$
(see, e.g., formula (61,2) in Ref.~\cite{LL-6}). Consequently, we get
\begin{equation}\nonumber
  \frac{d\eps}{dt}=\frac{d\eps}{dv}\cdot\frac{dv}{dt}
\end{equation}
on one hand and
\begin{equation}\nonumber
   \frac{d\eps}{dt}=\eps\frac{dl}{dt}=\eps\kappa v
\end{equation}
on the other hand, so that equality of these two expressions yields
\begin{equation}\label{eq24}
  \frac{dv}{dt}=\frac{v\eps}{d\eps/dv}\kappa=\frac{\eps}{m_s}\cdot\kappa,
\end{equation}
where $m_s=2d\eps/dv^2<0$ is an ``effective soliton mass'' per unit length. Now we
take into account the stretching of the contour with the local velocity $-w$ and
replace $dv/dt$ by the substantial derivative (\ref{eq22}):
\begin{equation}\label{eq26}
  v_t-\int_0^s\kappa vds'\cdot v_s=\frac{\eps}{m_s}\cdot\kappa.
\end{equation}
Equations (\ref{eq21}) and (\ref{eq26}) comprise the system of the contour dynamics
equations. For the case of the nonlinearity $f(\rho)=\rho$ they were derived in
Ref.~\cite{ms-10,mss-11} from the Gross-Pitaevskii equation (\ref{eq1}) by means of the
regular perturbation theory.

As a simple application of these equations, let us consider a linear approximation
when a straight soliton ($\kappa_0=0$) moving with velocity $v_0$ is slightly
disturbed and the above equations reduce to ($\kappa=\kappa_0+\kappa'$, $v=v_0+v'$)
\begin{equation}\label{eq27}
  \kappa'_t\approx v'_{yy},\qquad v'_t\approx \frac{\eps}{m_s}\cdot\kappa'.
\end{equation}
Looking for the solution in the form $\kappa',v'\propto\exp(iky+\Gamma t)$ we find
\begin{equation}\label{eq28}
  \Gamma=-\frac{\eps}{m_s}k^2,
\end{equation}
that is we have reproduced the result of Ref.~\cite{kp-08}.

Now we can turn to more interesting self-similar solutions of the obtained equations.

\section{Self-similar solution}

As was noticed in Ref.~\cite{ms-10,mss-11}, equations (\ref{eq21}) and (\ref{eq26}) 
are invariant with respect to the scaling transformation
$s=\al\tilde{s},t=\al\tilde{t},\kappa=\tilde{\kappa}/\al,v=\tilde{v}$. Therefore this
system has the solution in the form
\begin{equation}\label{eq29a}
  v=V(\zeta),\qquad \kappa=\frac{K(\zeta)}{t},\qquad \zeta=\frac{s}{t},
\end{equation}
where $\zeta$ is a self-similar variable. Such a form of the solution implies that
in the limit $t\to-0$ the solution becomes singular, that is the contour dynamics
approach loses its applicability when the radius of curvature $t/K$ becomes smaller
than the soliton's width. At the same time, the solution describes curved moving
solitons which can greatly deviate from their standard straight-line form.

Substitution of (\ref{eq29a}) into (\ref{eq21}) and (\ref{eq26}) yields
\begin{equation}\label{eq29}
\begin{split}
  & \frac{d(\zeta K)}{d\zeta}+\frac{d}{d\zeta}\left(K\int_0^{\zeta}VKd\zeta'\right)
  =-\frac{d^2V}{d\zeta^2},\\
  & \left(\zeta+\int_0^{\zeta}VKd\zeta'\right)\frac{dV}{d\zeta}=-\frac{\eps}{m_s}\cdot\kappa.
  \end{split}
\end{equation}
Following \cite{ms-10,mss-11}, we introduce the function
\begin{equation}\label{eq30}
  \Phi(\zeta)=\zeta+\int_0^{\zeta}VKd\zeta',\quad
  K=\frac1V\left(\frac{d\Phi}{d\zeta}-1\right),
\end{equation}
and integrate the first equation (\ref{eq29}) to obtain
\begin{equation}\label{eq31}
  K\Phi=-\frac{dV}{d\zeta}+A,
\end{equation}
where the integration constant $A$ is determined by the condition
\begin{equation}\label{eq32}
  A=\left.\frac{dV}{d\zeta}\right|_{\zeta=0}.
\end{equation}
The second equation (\ref{eq29}) and (\ref{eq31}) give the system of
ordinary differential equations
\begin{equation}\label{eq33}
  \frac{dV}{d\zeta}=\frac{(\eps/m_s)A}{\eps/m_s-\Phi^2},\qquad
  \frac{d\Phi}{d\zeta}=1-\frac{AV\Phi}{\eps/m_s-\Phi^2}.
\end{equation}
We suppose that at the origin $s=\zeta=0$ the soliton is black, that is
$v(0)=0$, but the gradient of velocity $A$ is not equal to zero here. Then the system
(\ref{eq33}) must be solved with the initial conditions
\begin{equation}\label{eq34}
  V(0)=0,\qquad \Phi(0)=0.
\end{equation}

\begin{figure}[t]
\begin{center}
\includegraphics[width=8cm]{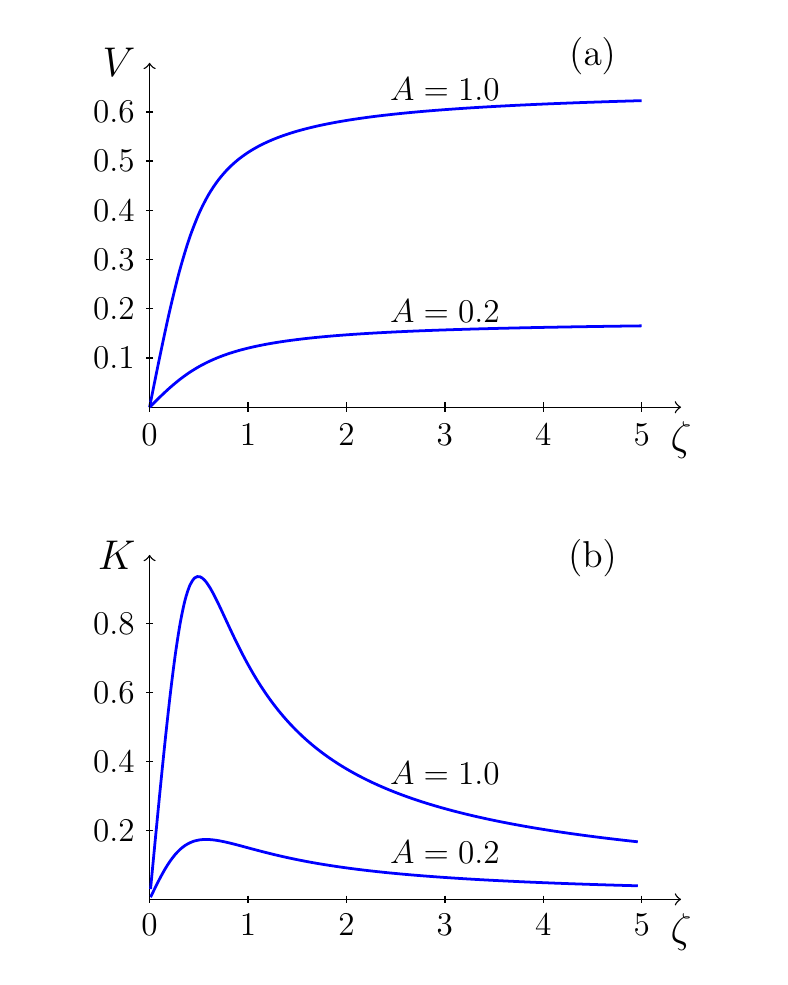}
\caption{
(a) Plot of the function $V(\zeta)$ for $f(\rho)=\rho$ and two values of the parameter $A$.
(b) Plot of the function $K(\zeta)$ for $f(\rho)=\rho$ and two values of the parameter $A$.
}
\label{fig1}
\end{center}
\end{figure}

For small $\zeta\ll1$ we get $\Phi\approx\zeta$, $V\approx A\zeta$,
$K\approx-\frac{A}{\eps/m_s}\zeta$, whereas for large $\zeta\gg1$ we obtain
the estimates $\Phi\sim\zeta$, $V\approx V_m=\mathrm{const}$,
$K\sim A/\zeta$. Consequently, the transition from one asymptotic regime
to the other one occurs at $\zeta\sim(|\eps/m_s|)^{1/2}$ and for small $A$
the solution has the order of magnitude $V\sim (|\eps/m_s|)^{1/2}A$ and
$K\sim A/(|\eps/m_s|)^{1/2}$. Hence, in case of small $A\ll1$ in the leading approximation
with respect to this small parameter we can put $V=0$ in the function $\eps/m_s$ and
consider this function as a constant parameter. Then the
first equation (\ref{eq33}) with $\Phi\approx\zeta$ becomes
\begin{equation}\label{eq35}
  \frac{dV}{d\zeta}=\frac{A}{1+\zeta^2/(\eps/|m_s|)}
\end{equation}
with the obvious solution
\begin{equation}\label{eq36}
  V(\zeta)=\sqrt{\frac{\eps}{|m_s|}}\,A\cdot
  \arctan\left(\frac{\zeta}{\sqrt{\eps/|m_s|}}\right).
\end{equation}
With the same accuracy we obtain from (\ref{eq31})
\begin{equation}\label{eq37}
  K(\zeta)=\frac{A\zeta}{\eps/|m_s|+\zeta^2}.
\end{equation}
In the limit $\zeta\to\infty$ we find
\begin{equation}\label{eq38}
  V\approx V_m=\frac{\pi}2\sqrt{\frac{\eps}{|m_s|}}\,A,\quad
  K\approx\frac{A}{\zeta},\quad A\ll1.
\end{equation}

\begin{figure}[t]
\begin{center}
\includegraphics[height=5cm]{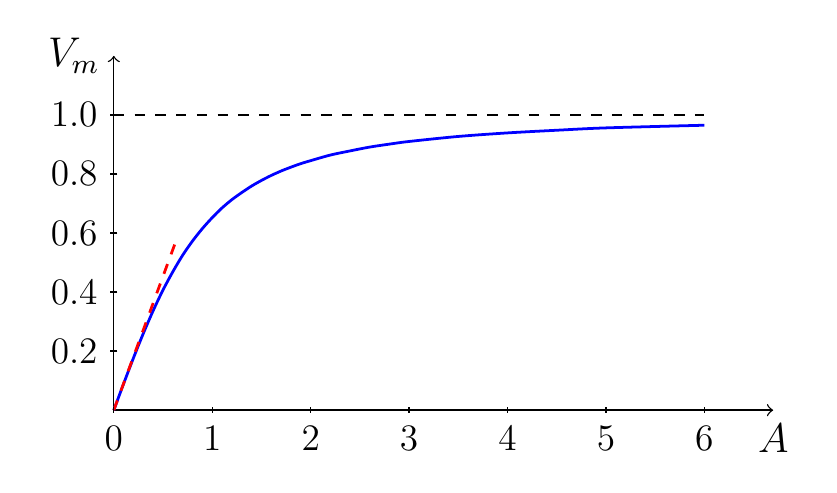}
\caption{
The dependence of the limiting velocity $V_m$ on the parameter $A$.
The red dashed line corresponds to Eq.~(\ref{eq38})
applicable for $A\ll1$.
}
\label{fig2}
\end{center}
\end{figure}

In case of large $A$ the system (\ref{eq33}) is to be solved numerically.
For example, if we take in Eq.~(\ref{eq1}) the Kerr-like nonlinearity
$f(\rho)=\rho$, then we get
\begin{equation}\label{eq39}
  \frac{\eps}{m_s}=-\frac13(1-V^2),
\end{equation}
where we have assumed $\rho_0=1$, and the system (\ref{eq33}) takes the form
\begin{equation}\label{eq40}
  \frac{dV}{d\zeta}=\frac{A(1-V^2)}{1-V^2+3\Phi^2},\quad
  \frac{d\Phi}{d\zeta}=1+\frac{3AV\Phi}{1-V^2+3\Phi^2}.
\end{equation}
Plots of its solutions for two different values of $A$ are depicted in
Fig.~\ref{fig1} (see also \cite{ms-10,mss-11}). These solutions confirm the above
estimates. The dependence of the limiting value $V_m$ on $A$ is shown in
Fig.~\ref{fig2}, where the red dashed line corresponds to the formula
$V_m\approx(\pi/2\sqrt{3})A$ which is a particular case of Eq.~(\ref{eq38})
for the Kerr-like nonlinearity. As we see, the agreement with the limit of
small $A$ is good enough for $A\lesssim0.5$. In the asymptotic region
$\zeta\gg1$ the first equation (\ref{eq40}) reduces to
\begin{equation}\label{eq41}
  \frac{dV}{d\zeta}=\frac{A(1-V^2)}{3\zeta^2}
\end{equation}
and it can be easily integrated to give
\begin{equation}\label{eq42}
  V(\zeta)\approx
  V_m-\frac{A(1+V_m)^2}{3\zeta},\qquad \zeta\gg1,
\end{equation}
where the integration constant is chosen according to the condition
$V(\zeta)\to V_m$ as $\zeta\to\infty$. This formula agrees with the
asymptotic expression
\begin{equation}\label{eq43}
  V(\zeta)\approx \frac{\pi}{2\sqrt{3}}\,A-\frac{A}{3\zeta},\qquad A\ll1,
\end{equation}
obtained from (\ref{eq36}) in the limit $\zeta\to\infty$.

\begin{figure}[t]
\begin{center}
\includegraphics[height=8cm]{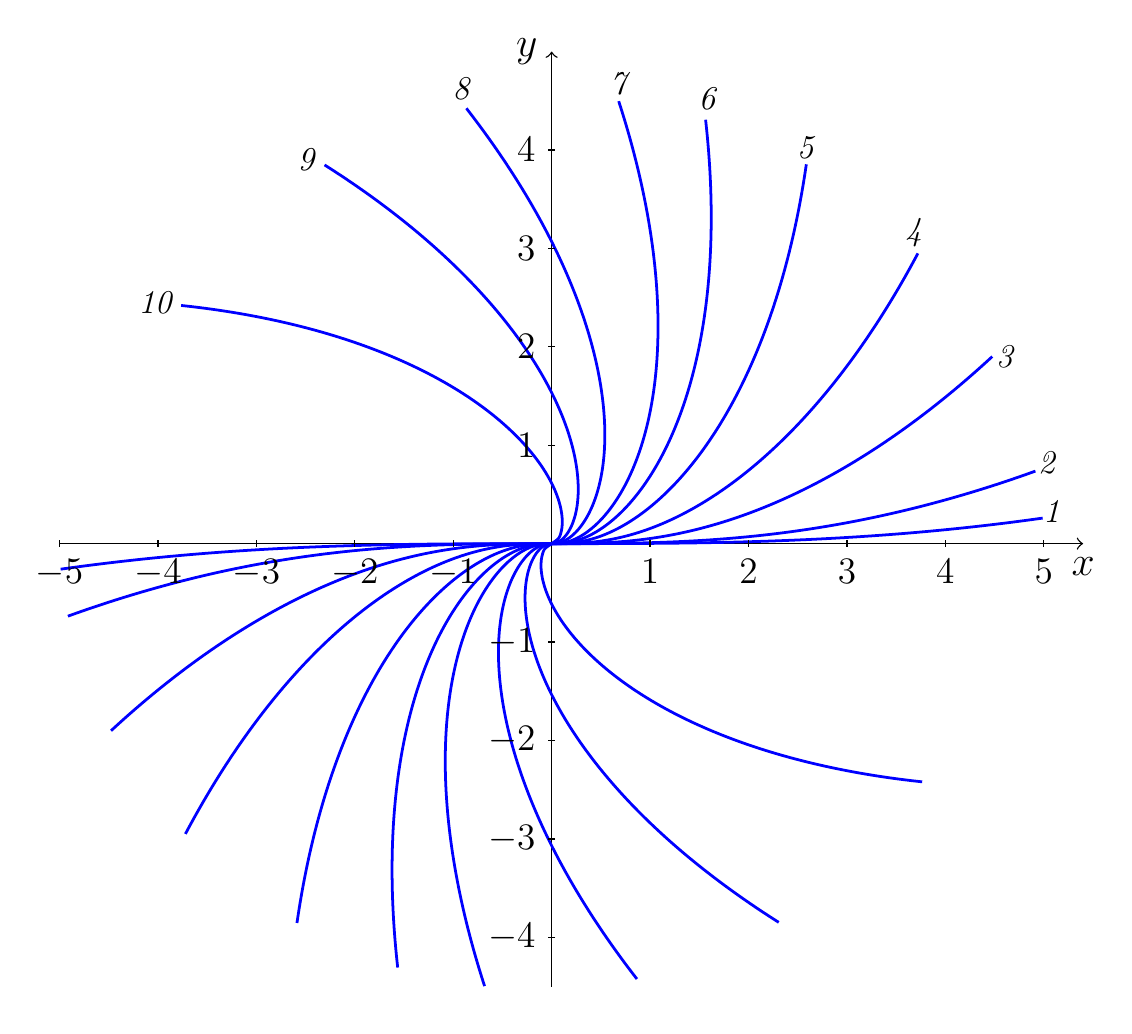}
\caption{
Dark soliton contours at different moments of time: (\emph{1}) $t=-10$;
(\emph{2}) $t=-5$; (\emph{3}) $t=-2$; (\emph{4}) $t=-1$; (\emph{5}) $t=-0.5$;
(\emph{6}) $t=-0.3$; (\emph{7}) $t=-0.2$; (\emph{8}) $t=-0.1$; (\emph{9}) $t=-0.05$;
(\emph{10}) $t=-0.02$.
All of the curves correspond to $A=0.5$.
}
\label{fig3}
\end{center}
\end{figure}

To find the form of the soliton at the moment $t$, we choose for definiteness
$(x,y)$-coordinates in such a way that the tangent vector $\bt$ can be
written in the form
\begin{equation}\label{eq44}
  \bt=(\cos\theta,\sin\theta)\quad \text{or}\quad
  \frac{dx}{ds}=\cos\theta,\quad \frac{dy}{ds}=\sin\theta,
\end{equation}
and $\theta=0$ at $s=0$. Then from the first equation (\ref{eq17}) we find
at once
\begin{equation}\label{eq45}
  \kappa=\left|\frac{d\bt}{ds}\right|=\frac{d\theta}{ds}.
\end{equation}
In case of small $A$ we obtain with the use of (\ref{eq37}) the expression
for the curvature,
\begin{equation}\label{eq46}
  \kappa=\frac{K}t=\frac{As}{(\eps/m_s)t^2+s^2}.
\end{equation}
Consequently, integration of Eq.~(\ref{eq45}) gives
\begin{equation}\label{eq47}
  \theta(s)=\frac{A}2\ln\left[1+\frac{s^2}{(\eps/|m_s|)t^2}\right].
\end{equation}
At last, integration of Eqs.~(\ref{eq44}) yields the soliton's contour in
a parametric form,
\begin{equation}\label{eq48}
\begin{split}
  &x(s,t)=\int_0^s\cos\left\{\frac{A}2\ln\left[1+\frac{s^2}{(\eps/|m_s|)t^2}\right]\right\}ds,\\
  &y(s,t)=\int_0^s\sin\left\{\frac{A}2\ln\left[1+\frac{s^2}{(\eps/|m_s|)t^2}\right]\right\}ds.
  \end{split}
\end{equation}
The integrals here can be expressed in terms of the hypergeometric function
(see, e.g., \cite{ww-27})
\begin{equation}\label{eq49}
  x(s,t)+iy(s,t)=sF\left(\frac12,-\frac{iA}2,\frac32,-\frac{s^2}{(\eps/|m_s|)t^2}\right).
\end{equation}
For small $|s|\ll|t|$ we get
\begin{equation}\label{eq50}
  x(s,t)\approx s,\qquad y(s,t)\approx\frac{As^3}{6(\eps/|m_s|)t^2},
\end{equation}
that is the soliton has the form of a cubic parabola here,
\begin{equation}\label{eq51}
   y(x,t)\approx\frac{Ax^3}{6(\eps/|m_s|)t^2},\qquad |x|\ll|t|.
\end{equation}
The entire contour has the form of a spiral shown in Fig.~\ref{fig3}
for different values of $t$. These curves have the maximal curvature
\begin{equation}\label{eq52}
  |\kappa_{\text{max}}|=\frac{\sqrt{\eps/|m_s|}A}{2|t|}
\end{equation}
at $s=\pm\sqrt{\eps/|m_s|}\,t$ and the coordinates of the points with the
maximal curvature are to be found from the equation
\begin{equation}\label{eq53}
  x+iy=\pm \sqrt{\frac{\eps}{|m_s|}}\,t  F\left(\frac12,-\frac{iA}2,\frac32,-1\right).
\end{equation}
The minimal radius of the curvature $1/|\kappa_{max}|\propto|t|$ decreases as
$t\to-0$ and when it reaches the order of magnitude of the soliton's width,
the contour dynamics approach loses its applicability. Numerical solution of the
Gross-Pitaevskii equation performed in Refs.~\cite{ms-10,mss-11,kk-11} shows that at this stage
of evolution the dark soliton breaks with formation of vortex-antivortex pairs.
The developed here theory describes the soliton's evolution before this breaking moment.

\section{Conclusion}

We developed further the method of contour dynamics of dark solitons suggested first
in Ref.~\cite{ms-10,mss-11}. A simplified derivation of equations of contour dynamics
is given for the general form of the nonlinearity function in the
Gross-Pitaevskii equation. The self-similar solution of the obtained equations
is studied in detail. The results of this paper provide estimates for typical
characteristics of dark solitons and the time of their breaking to vortex-antivortex
pairs. We have considered evolution of solitons evolving in a uniform quiescent
background, but the simple method used here can be applied to situations with
non-uniform flowing condensates.

\begin{acknowledgments}
I am grateful to V.~A.~Mironov, A.~I.~Smirnov, and L.~A.~Smirnov for useful
discussions of their remarkable papers \cite{ms-10,mss-11}. I also thank
Y.~A.~Stepanyants for important remarks.
The reported study was partially funded by RFBR, project number 20-01-00063.
\end{acknowledgments}


\begin{thebibliography}{90}%

\bibitem{ka-03} Yu.~S.~Kivshar,  G.~P.~Agraval, {\it Optical solitons.
From Fibers to Photonic Crystals,} (Academic Press, N. Y., 2003).

\bibitem{ps-03} L.~Pitaevskii,  S. Stringari, {\it Bose-Einstein Condensation,}
(Clarendon Press, Oxford, 2003).

\bibitem{ba-00} Th.~Busch, J.~R.~Anglin, Phys. Rev. Lett. {\bf 84,} 2298 (2000).

\bibitem{kp-04}  V.~V.~Konotop, L.~P.~Pitaevskii, Phys. Rev. Lett. {\bf 93,} 240403 (2004).

\bibitem{bkp-06} V.~A.~Brazhnyi, V.~V.~Konotop, L.~P.~Pitaevskii,
Phys. Rev. A {\bf 73,} 053601 (2006).

\bibitem{ks-09} A.~M.~Kamchatnov, M.~Salerno,
J. Phys. B: At. Mol. Opt. Phys. {\bf 42,} 185303 (2009).

\bibitem{dngl-13} B.~Deveaud, G.~Nardin, G.~Grosso, Y.~L\'{e}ger,
Dynamics of Vortices and Dark Solitons in Polariton Superfluids, in
{\it Physics of Quantum Fluids,} Eds. A.~Bramati, M.~Modugno, p.~99, (Springer, Heidelberg, 2013).

\bibitem{kp-70} B.~B.~Kadomtsev, V.~I.~Petviashvili, 
Sov. Phys. Dokl. {\bf 15,} 539 (1970).

\bibitem{zakh-75} V.~E.~Zakharov, JETP Lett. {\bf 22,} 172 (1975).

\bibitem{kt-88} E.~A.~Kuznetsov, S.~K.~Turitsyn, Sov. Phys. JETP {\bf 67,} 1583 (1988).

\bibitem{tcldk-96} V.~Tikhonenko, J.~Christou, B.~Luther-Davies, Y.~S.~Kivshar,
Opt. Lett. {\bf 21,} 1129 (1996).

\bibitem{msz-96} A.~V.~Mamaev, M.~Saffman, A.~A.~Zozulya, Phys. Rev. Lett. {\bf 76,} 2262 (1996).

\bibitem{msaz-96} A.~V.~Mamaev, M.~Saffman, D.~Z.~Anderson, A.~A.~Zozulya,
Phys. Rev. A {\bf 54,} 870 (1996).

\bibitem{ahrfccc-01} B.~P.~Anderson, P.~C.~Haljan, C.~A.~Regal, D.~L.~Feder, L.~A.~Collins,
C.~W.~Clark, E.~A.~Cornell, Phys. Rev. Lett. {\bf 86,} 2926 (2001).

\bibitem{psk-95} D.~E.~Pelinovsky, Y.~A.~Stepanyants, Y.~S.~Kivshar, Phys. Rev. E {\bf 51,} 5016 (1995).

\bibitem{kp-2000} Y.~S.~Kivshar, D.~E.~Pelinovsky, Phys. Rep. {\bf 331,} 117 (2000).

\bibitem{ms-10} V.~A.~Mironov, L.~A.~Smirnov, 
Bull. Russian Acad. Sci. Phys., {\bf 74,} 1699 (2010). 

\bibitem{mss-11} V.~A.~Mironov, A.~I.~Smirnov, L.~A.~Smirnov, JETP {\bf 112,} 46 (2011).

\bibitem{bkkl-84} R.~C.~Brower, D.~A.~Kessler, J.~Koplik, H.~Levine, Phys. Rev. A {\bf 29,} 1335 (1984).

\bibitem{ssok-14} L.~A.~Smirnov, D.~A.~Smirnova, E.~A.~Ostrovskaya, Y.~S.~Kivshar,
Phys. Rev. B {\bf 89,} 235310 (2014).

\bibitem{zhr-79} N.~J.~Zabusky, M.~H.~Hughes, K.~V.~Roberts,   
J. Comput. Phys. {\bf 30,} 96 (1979).

\bibitem{pullin-92} D. L Pullin, Annu. Rev. Fluid Mech. {\bf 24,} 89 (1992).

\bibitem{pressley} A.~Pressley, {\it Elementary Differential Geometry,} (Springer, London, 2010).

\bibitem{os-76} L.~A.~Ostrovsky, V.~I.~Shrira, Sov. Phys. JETP {\bf 44,} 738 (1976).

\bibitem{kp-08} A.~M.~Kamchatnov, L.~P.~Pitaevskii, Phys. Rev. Lett. {\bf 100,} 160402 (2008).

\bibitem{LL-6} L.~D.~Landau, E.~M.~Lifshitz, {\it Fluid Mechanics,}
(Pergamon Press, Oxford, 1987).

\bibitem{ww-27} E. T. Whittaker, G. N. Watson, {\it A Course of Modern Analysis,}
(CUP, Cambridge, 1927).

\bibitem{kk-11} A.~M.~Kamchatnov, S.~V.~Korneev, Phys. Lett. A {\bf 375,} 2577 (2011).



\end{thebibliography}
\end{document}